# Tunnel magnetoresistance in scandium nitride magnetic tunnel junctions using first principles

**Authors**: Suyogya Karki[1] (suyogya.karki@utexas.edu), Vivian Rogers[1] (vivian.rogers@utexas.edu), Priyamvada Jadaun[1] (priyamvada@utexas.edu), Daniel S. Marshall[2-3] (DMarshall@tae.com), and Jean Anne C. Incorvia*[1] (incorvia@austin.utexas.edu)

[1]UT Austin Electrical & Computer Engineering Department, 2501 Speedway, EER 3.820, Austin, TX 78712

[2]TAE Technologies, Inc., 19631 Pauling, Foothill Ranch, CA 92610

[3]Arizona State University, SEMTE Dept., 501 E. Tyler Mall, ERC 301, Tempe, AZ 85281

* Corresponding author

**Abstract**

The magnetic tunnel junction is a cornerstone of spintronic devices and circuits, providing the main way to convert between magnetic and electrical information. In state-of-the-art magnetic tunnel junctions, magnesium oxide is used as the tunnel barrier between magnetic electrodes, providing a uniquely large tunnel magnetoresistance at room temperature. However, the wide bandgap and band alignment of magnesium oxide-iron systems increases the resistance-area product and causes challenges of device-to-device variability and tunnel barrier degradation under high current. Here, we study using first principles narrower-bandgap scandium nitride tunneling properties and transport in magnetic tunnel junctions in comparison to magnesium oxide. These simulations demonstrate a high tunnel magnetoresistance in Fe/ScN/Fe MTJs via $\Delta_1$ and $\Delta_2'$ symmetry filtering with low wavefunction decay rates, allowing a low resistance-area product. The results show that scandium nitride could be a new tunnel barrier material for magnetic tunnel junction devices to overcome variability and current-injection challenges.

**Introduction**

Magnetic tunnel junctions (MTJs) are basic building blocks for emerging spintronic devices, including for spin transfer torque magnetic random access memory (STT-MRAM)[1], a leading emerging nonvolatile memory that is steadily transitioning into production, as well as for magnetic logic-in-memory[2-4] and neuromorphic computing[5,6] applications.

An MTJ consists of a thin insulating tunnel barrier sandwiched between two ferromagnetic (FM) electrodes. When a current is passed across the barrier, parallel ($P$) magnetization of the two electrodes provides a higher density of states for the majority spin electrons to tunnel across, giving a low resistance, $R_P$, state. When the magnetization of the two electrodes is antiparallel ($AP$), the device is in a high resistance, $R_{AP}$, state. These two states can be used as 1's and 0's for memory and logic applications. The MTJ is characterized by the tunnel magnetoresistance $TMR = \frac{R_{AP} - R_P}{R_P} \times 100\%$ and the resistance-area product $RA = R_P \times A$, where $A$ is the cross-sectional area of the junction.

Progress in MTJ devices has been driven by materials revolutions. Spin-dependent tunneling behavior was first implemented in amorphous aluminum oxide ($Al_2O_3$) tunnel junctions[7], which have shown a room temperature $TMR$ up to ~70%. About ten years later, a larger $TMR$ was theoretically predicted and then measured in magnesium oxide (MgO) barriers[8], showing $TMR$ above 600%[9] at room temperature in experiments. MgO has been the tunnel barrier of choice for over ten years because of its unique spin transport properties: MgO(001) can lattice match to Fe(001) to prevent mixing of electron states as they tunnel across the barrier, and selectively filters out all tunneling symmetries except the $\Delta_1$ band in Fe. Fe's $\Delta_1$ band is highly spin-polarized and has no available states at the Fermi level in the minority spin, which allows for a theoretically ultra-high magnetization-dependent tunneling via Bloch waves with small transverse momentum. In the minority spin channel, the smaller conductance is mainly due to interface resonance states[10].

However, there are challenges to using MgO that hinder MTJs and their associated technologies from competing with other emerging memories[11]. A major challenge is that very thin tunnel barriers are necessary. The MgO layer is grown < 1.5 nm thick to have a reasonably low $RA$ product. This is due to both the wide 7.8 eV bandgap of MgO and the metal-insulator band alignment. Pinholes present in this thin layer can create a path for current and degrade the $TMR$, and when so thin, slight variations of the thickness across a wafer create variations in the $TMR$ and $RA$ product[12,13]. In addition, an FeO interlayer is created upon annealing at higher temperatures, decreasing expected $TMR$ values in experiment[14]. This makes MgO MTJs a challenge to grow, especially compared to competing technologies such as resistive random-access memory. It is also a hindrance for advanced applications of MTJ devices in artificial intelligence and neuromorphic computing, where current may be applied across the tunnel barrier often during real-time adaptation to the environment and where novel switching methods may be used that require higher current across the barrier, causing device breakdown[15,16]. These challenges lend credence to investigating alternative materials to MgO that can have similar symmetry-filtering transport properties with a narrower bandgap, although no materials have been able to compete with the $TMR$ of MgO to-date at room temperature.

Here, we study using first principles the tunneling properties of scandium nitride (ScN) to understand the material's transport characteristics and determine if MTJs using ScN can achieve high $TMR$. ScN is chosen for this study because it has a narrower bandgap than MgO (indirect transition from the X to Γ point of 0.9-1.6 eV and direct transition at the X point of 2-2.9 eV[17]), and it has a similar rock salt crystal structure to MgO. ScN(001) is face-centered cubic (FCC) with lattice constant $a$ = 4.501 Å[17], compared to MgO(001), also FCC with $a$ = 4.212 Å[18]. ScN is a group IIIB transition metal nitride and has not been widely used in device applications[19], but has found niche uses in GaN crystal growth[20,21] and radio frequency devices[22].

There is ongoing research in developing barrier materials with large $TMR$ ratios to try to compete with MgO, largely in the class of Mg oxide alloys[23-25]. In one simulation study, ZnO[26] with rock salt structure, bandgap 2.132 eV at the Γ point, and indirect gap of 0.913 eV showed $TMR$ = 446% and $RA$ = 0.0468 Ω-$\mu m^2$. There is also recent interest in exploring alternate electrode materials for higher spin polarization; one such example is the use of Heusler compounds for both electrodes and barrier materials. Recent work on NiMnSb-MgO junctions[27] predicted high $TMR > 10^4$. To our knowledge, no work has investigated nitride-based tunnel barriers, nor specifically ScN MTJs.

**Results**

*Materials Structure*

We investigate the complex band structure of ScN, as well as electron transport in Fe/ScN/Fe MTJs, to understand ScN's properties as a tunnel barrier using density functional theory (DFT) and plane wave conductance techniques. Figure 1a shows the converged ScN lattice with a rock salt crystal structure, and Fig. 1b shows an example converged Fe/ScN/Fe supercell using visualization software VESTA[28]. Supercells of both Fe/ScN(001)/Fe and Fe/MgO(001)/Fe systems are created for $t$ = 3-7 atomic layers of the barrier region to compare the ScN and MgO behavior. In $x$ and $y$, the unit cell is repeated to infinity; in $z$ the Fe leads extend to infinity in both directions.

To interface with Fe electrodes, the ScN and MgO unit cells are rotated by 45° around the $z$ direction such that the anion atoms (O or N) in the barrier region are positioned directly above the Fe at the interface. After this rotation, the lattice parameter is constrained to the experimental lattice constant of Fe (2.866 Å) in the $x$ and $y$ directions. This induces a 3.9% in-plane compressive strain in the barrier to match the experimental lattice parameter of MgO (4.212 Å), and for ScN the experimental lattice parameter of 4.501 Å is compressed by 11% to match with the leads. These supercells are relaxed using the Vienna ab-initio simulation program (VASP)[29-31] with molecular dynamics. The wavefunctions and resulting conductance of the system are calculated using the Quantum Espresso package[32]. Magnetization of the Fe electrodes are collinear with the $z$ axis.

The spin up ($T^\uparrow$) and spin down ($T^\downarrow$) transmission probabilities, i.e. the probability for transmitting an electron that enters the channel, is calculated. The conductance is then calculated using the Landauer formula $G^\uparrow = \frac{e^2}{h}T^\uparrow$ and $G^\downarrow = \frac{e^2}{h}T^\downarrow$, where $e$ is electron charge and $h$ is Planck's constant for both spin up (majority) and spin down (minority) channels. The Fe electrodes are set up in the $P$ state magnetized in +$z$; for the $AP$ state, the bottom Fe electrode remains in +$z$ and the top is in -$z$. The total conductance ($G$) for each magnetization state of the electrodes is calculated by adding the conductance from both the spin up and spin down channel, $G = G^\uparrow + G^\downarrow$.

The resulting converged real band structure is shown in Fig. 2 for bulk ScN without and with a Hubbard potential (U) of 4.5 eV added to the 3d orbital of Sc. DFT simulations usually underestimate bandgaps, and the value depends on the pseudopotential used for the calculation. Using the Perdew-Burke-Ernzerhof (PBE) pseudopotential in Fig. 2a, an indirect gap of 0 eV is observed in the band structure. With the additional Hubbard potential, in Fig. 2b there is a direct gap at the Γ point of 2.99 eV, a direct gap at the X point of 2.28 eV, and an indirect Γ-X gap of 1.31 eV. These results are in agreement with a previous DFT+U simulation, which showed the ScN band structure having a direct gap at the Γ point of 3.39 eV and a direct gap at the X point of 1.55 eV[33]. The calculation is also comparable to experimental values, where gaps are observed at the Γ point of 3.8 eV, at the X point of 2.4 eV, and an indirect Γ-X gap of 1.3 ± 0.3 eV[34].

Figure 3 shows the complex band structures of ScN (Fig. 3a) and MgO (Fig. 3b) sampled at the Γ and X points. Where the complex band energies intersect with the Fermi energy ($E = 0$) shows the rate of decay of that band in the barrier material. While the band diagram used for conduction calculations agrees with the experimental bandgap, for the complex band calculation of ScN the bandgap is smaller than seen in experiment, which is acceptable for understanding which bands are contributing to conduction (see Supplementary Methods).

For MgO, Fig. 3b shows Im(k) = 0.21 $\frac{2\pi}{a}$ at the Γ point, showing MgO's $\Delta_1$ band has a low decay rate as expected[10], such that the majority electrons from Fe will continue through the barrier. For ScN, Fig. 3a shows a low decay rate of Im(k) = 0.07 $\frac{2\pi}{a}$ at the Γ point, as well as a low decay rate of Im(k) = 0.07 $\frac{2\pi}{a}$ at the X point, showing both $\Delta_1$ and $\Delta_2'$ are expected to contribute to tunneling. Since $\Delta_2'$ is also spin polarized in Fe (see Supplementary Fig. 1), these results show that ScN junctions are expected to have spin-dependent tunneling, but with contribution from both of these bands, differing from MgO. We also note that even with the narrower bandgap than experiment, we can conclude that the $\Delta_1$ decay rate in ScN is smaller than in MgO, providing higher majority channel conduction than MgO.

*Spin-Dependent Transport*

In Fig. 4 the transmission probability $T$ for Fe/ScN($t$ = 6)/Fe MTJs (top row) is compared to Fe/MgO(6)/Fe MTJs (bottom row) for both spin channels and both magnetic orientations of the Fe electrodes. The plots show the transmission probability for each $k_x$ and $k_y$ point in the supercell Brillouin zone centered at the $\Gamma$ point, using a 200 x 200 $k$-grid. In the transmission plot for the majority $P$ channel of the ScN MTJ, Fig. 4a, we see that conduction occurs at the center of the Brillouin zone ($\Gamma$ point) and along the X and M points of the tetragonal supercell. The conduction band minimum in the ScN band structure is at the X point and the ScN lattice has been rotated 45° around the $z$ axis for the supercell. As expected, this is reflected in the high transmission lobes around the supercell M points. These peaks corresponding to bulk ScN's X point suggest that the spin-polarized $\Delta_2'$ symmetries of the electrodes play a large role in tunneling through ScN junctions, important for optimizing the electrode material for ScN MTJs.

In comparison, the MgO majority $P$ channel, Fig. 4e, shows a broad peak centered at the $\Gamma$ point via the Fe $\Delta_1$ band. Integrating the majority channel transmission over the Brillouin zone, we find that the total $T_P^\uparrow = 6.44 \times 10^{-3}$ for 6 layers of ScN and $T_P^\uparrow = 1.74 \times 10^{-4}$ for 6 layers of MgO. Figures 4b and 4f show transmission plots of the minority channel for ScN and MgO when the electrode magnetization is parallel. In the minority channel for ScN, $T_P^\downarrow = 8.63 \times 10^{-5}$ and for MgO $T_P^\downarrow = 2.80 \times 10^{-5}$. For the ScN minority $P$ channel we see a peak in transmission at the $\Gamma$ and X points, whereas the minority $P$ channel of MgO shows transmission around the X points.

Figures 4c-4d show the conduction through ScN when the electrodes have $AP$ magnetization, compared to MgO $AP$ in Figs. 4g-4h. Compared to Figs. 4a-4b, the transmission in the spin up and spin down channels in the ScN system show a much broader peak around the $\Gamma$ point with lobes

around the X point. In MgO, the transmission peaks for both channels (Fig. 4g and Fig. 4f) are primarily seen around the Γ point, with most transmission around the X point filtered out compared to the *P* minority channel. In the ScN system, the total spin up transmission in *AP* configuration $T_{AP}^{\uparrow} = 2.89 \times 10^{-5}$ and in MgO $T_{AP}^{\uparrow} = 3.25 \times 10^{-7}$. The spin down channels for both ScN and MgO supercells with *AP* magnetization show similar transmission values to that of the spin up channels. The results for Fe/MgO/Fe agree with previous work[10] and give confidence to the validity of the model. As expected, MgO shows high spin filtering: the majority channel dominates conductance compared to the minority channel, and thus parallel magnetized electrodes show a greater than 300× increase in conductance compared to antiparallel magnetized electrodes for the 6-layer system. These essential features lead to the high $TMR$ seen in MgO MTJs. Promisingly, ScN shows similar spin filtering properties, where, for this layer number, the majority channel in the parallel magnetized electrodes shows a 75× higher conductance compared to the minority channel. Also, the conductance is 120× higher in the parallel magnetized electrodes compared to the antiparallel electrodes, indicating that ScN MTJs can also achieve high $TMR$.

*Magnetic Tunnel Junction Properties*

Using the resulting transmission probability values, the conductance ($\frac{e^2}{h}$) for majority and minority channels for *P* and *AP* orientation of the electrodes is shown in Table 1 for ScN and Table 2 for MgO. The total conductance ($G$) in units of siemens ($\Omega^{-1}$) for each *P* and *AP* configuration is then calculated by summing $G^{\uparrow}$ and $G^{\downarrow}$. The conductance is converted to $TMR$ for the various barrier thicknesses simulated. In our simulations, the $t = 4$ layered system had unstable molecular dynamics convergence and thus was not calculated (see Supplementary Methods). It is seen that in the ScN systems, $TMR = $ 108-11,800% showing that ScN can demonstrate large TMR. In comparison, the Fe/MgO/Fe system $TMR$ = 408-41,800%, in agreement with previous simulation results[10]. Although simulations predict ultra-high values of $TMR$ for both systems, it is difficult to achieve this in experiments. Our model validates that spin-dependent tunneling is achieved in the ScN MTJs with additional tunneling mechanisms compared to MgO MTJs.

It is observed in Table 1 for Fe/ScN/Fe MTJs there is a non-exponential decay of conductance in each channel with respect to barrier thickness, motivating an explanation for the deviation from the expected relationship seen in MgO MTJs. The breakdown of crystallinity due to lattice mismatch of ScN and Fe electrodes and the unique $\Delta_2'$ conduction through the X point could explain this behavior. For the majority channel in MgO, most of the conduction is due to tunneling states with little transverse momentum at the $\Gamma$ point. In ScN, there are additional tunneling states around the X points, which have electron plane waves of a higher spatial frequency than those around the $\Gamma$ point. This suggests strong layer-dependent wavefunction interference within the barrier region, as seen in the interface resonance in MgO's minority spin electron transport[10].

In the molecular dynamics simulation, the ScN lattice is compressed in *x* and *y* to match the Fe electrodes, buckling the barrier layer slightly and decreasing the crystallinity of the ScN. The simulations show that this atomic convergence varies with respect to barrier layer number. Figures 5a-c show the converged Fe/ScN/Fe structure for *t* = 3, 6, and 7 atomic layers of ScN. Figure 5d quantifies the deviation $\Delta z_i = z_{Sc} - z_N$ for every barrier layer in each converged MTJ supercell. For the 6 and 8 layer systems, the offset $\Delta z_i$ is centered at zero, preserving the mirror symmetry in *z* that contributes to high $TMR$. This can explain why the $TMR$ is lower in the 3, 5, and 7 layer simulations than the 6 and 8 layer simulations. In a real device, we expect the Fe/ScN interface to more likely follow the 6 and 8 layer systems' behavior, since the ScN lattice will be allowed to relax.

Figure 6 compares the $RA$ product between ScN and MgO MTJs for varying barrier layer number $t$. The cross-sectional area of the junction used for the calculation is $8.21 \times 10^{-8}$ $\mu m^2$. In MgO systems, as $t$ increases, $RA$ increases from 0.203 to 51.0 $\Omega-\mu m^2$; e.g. for MgO $t$ = 6 layers, $RA$ = 10.5 $\Omega-\mu m^2$ and $TMR$ = 30,000%. In ScN MTJs, $RA$ ranges between 0.227 and 7.31 $\Omega-\mu m^2$ and stays low with increasing $t$; e.g. for ScN $t$ = 6 layers, $RA$ = 0.326 $\Omega-\mu m^2$ and $TMR$ = 11,200%. The small decay rate of both $\Delta_1$ and $\Delta_2'$ channels, combined with the narrower bandgap of ScN, can explain why the majority channel is highly conductive at all thicknesses simulated.

Calculating the power $W = \frac{I^2 \times RA}{A}$, the power consumed for the $t$ = 6 ScN device is 3.11% of that consumed by the MgO device with current held constant. Alternatively, if $W$ is held constant, a 468% increase in the injected current for the ScN MTJ is seen compared to MgO. These results indicate that we expect to see a high $TMR$ in ScN systems with a low $RA$, enabling either higher current across thin tunnel barriers or thicker tunnel barriers while maintaining the same current, both of which could alleviate the variability and advanced application challenges facing MTJs today.

**Discussion**

We have shown from first principles that Fe/ScN/Fe MTJs have many of the desirable spin-dependent transport properties of Fe/MgO/Fe junctions, namely a low $\Delta_1$ decay rate through the $\Gamma$ point, but with additional tunneling pockets through the $\Delta_2'$ symmetries at the X point. We have shown this leads to a high $TMR$, competitive with that in Fe/MgO/Fe. The ScN MTJs also show a low *RA* product compared to MgO MTJs, which could be used to overcome variability and current-injection challenges in MgO MTJ devices. These results indicate that ScN could be an exciting new material for MTJ devices, in a field where few alternative materials to MgO have been developed. The work motivates experimental studies, exploration of other lower bandgap materials for advanced MTJ applications, investigation of effects of thermionic emission, and understanding of how the nitrogen-based barrier can affect device properties compared to traditional oxygen-based tunnel barriers.

**Methods**

*VASP*

The MTJ supercell was created for various layer numbers (*t*) for both ScN and MgO systems. The systems with odd *t* have 6 Fe atoms on the bottom side and 5 Fe atoms on the top side of the supercell. For the *t* = 6 systems, 6 atoms of Fe were placed on each side of the barrier to maintain the periodicity of the system and to ensure a match with the leads in conduction. The supercells were relaxed using the Vienna ab-initio simulation program (VASP) with molecular dynamics such that all forces are < 0.01 eV/Å. The convergence was also done for the total free energy and the band structure energy so that the change between two timesteps is < $10^{-5}$ eV. Using Perdew-Burke-Ernzerhof (PBE) functionals[35], the Fe/ScN/Fe and Fe/MgO/Fe supercells were converged using these cutoffs with an 11×11×11 *k* point mesh. The converged supercell atomic positions were then copied into Quantum Espresso to generate wavefunctions of the supercells and perform transport calculations.

*Quantum Espresso*

To generate the wavefunctions and calculate the complex bands and unit cell conductance, PWscf and PWcond were used from the Quantum Espresso package. The complex bands were generated with a 4-atom tetragonal unit cell for both bulk ScN and MgO using USPP PBE scalar-relativistic pseudopotentials, with the symmetry-resolved real bands sampled in the direction of conduction. Cutoffs of 50 Ry and 500 Ry were used for the wavefunction and charge density cutoffs. For ScN, a Hubbard +U of 8 eV was applied with the 'pseudo' projection method, though this was unable to fully correct the bandgap to the values used in the real band structure or MTJ conduction calculations.

After minimizing the energy of the MTJ unit cells in VASP, a PWscf self-consistency field (SCF) calculation was run using Scalar-relativistic PBE functionals with projector augmented wave potentials to generate the wave functions[36]. For all supercells, 64 Ry and 782 Ry were used for the wavefunction and charge density cutoffs, respectively. Marzari-Vanderbilt smearing[37] was used with a broadening parameter of 0.02. For the SCF calculation, an 11×11×1 Monkhorst-Pack *k*-grid[38] was used. For the systems with parallel magnetization of the leads, both leads were magnetized in the +$z$ direction, with the average magnetization of the Fe atoms converging to roughly 2.3 $\mu_B$. For the systems with antiparallel lead magnetization, the atomic positions of the parallel alignment system were duplicated in the +$z$ direction to create an Fe/Barrier/Fe-Fe/Barrier/Fe supercell. The far bottom and top Fe regions were magnetized in +$z$, and the middle Fe region was magnetized in -$z$, where the supercell would be cut in half for conduction calculations. This avoided modeling a discontinuity in magnetization to ensure that the Fe atoms in the barrier supercell matched up to the semi-infinite leads as they would in bulk Fe. The wavefunctions of the electrode unit cells were generated using the same parameters, except an 11×11×11 *k*-grid was used to reflect the cubic nature of the Fe unit cells in the bulk material.

In the SCF stage for the ScN systems, a DFT+U Hubbard offset of 4.5 eV was applied to the Sc atoms via the atomic projection method to match the bandgap with experimental values. We found this value by sweeping the Hubbard offset from 3 to 6 eV for bulk ScN and examining the band structure for each offset. Conductance and *k*-grid-resolved transmission mapping of each system were calculated with PWCOND under the Landauer-Büttiker[39] formalism. With each structure, the barrier supercell (half of the supercell for the antiparallel case) was interfaced to the semi-infinite Fe leads with appropriate magnetization direction. An energy window of 8 Ry was used for reducing the 2D plane wave basis set in transmission for the smaller systems, though the 7-layer systems were reduced to 6 Ry to improve stability. Both were converged to an accuracy of $10^{-8}$ Ry. The transmission was resolved with a 200×200 *k*-grid in *x* and *y*; a finely grained *k*-grid proved important for accurately capturing fine spikes in transmission.

**Data Availability**

The datasets generated during and/or analyzed during the current study are available from the corresponding author on reasonable request.

**Acknowledgements**

The authors acknowledge computing resources from the Texas Advanced Computing Center (TACC) at the University of Texas at Austin (http://www.tacc.utexas.edu), funding and discussions from Sandia National Laboratories, and funding from the Center for Dynamics and Control of Materials (CDCM) supported by the National Science Foundation under NSF Award Number DMR-1720595.

**Author Contributions**

S. Karki and V. Rogers carried out the simulations with assistance from P. Jadaun and prepared the manuscript. D. Marshall provided ideas and discussions of importance to the work. J. A. C. Incorvia led and supervised the work and manuscript.

**Competing Interests**

The authors declare no competing interests.

**Figures**

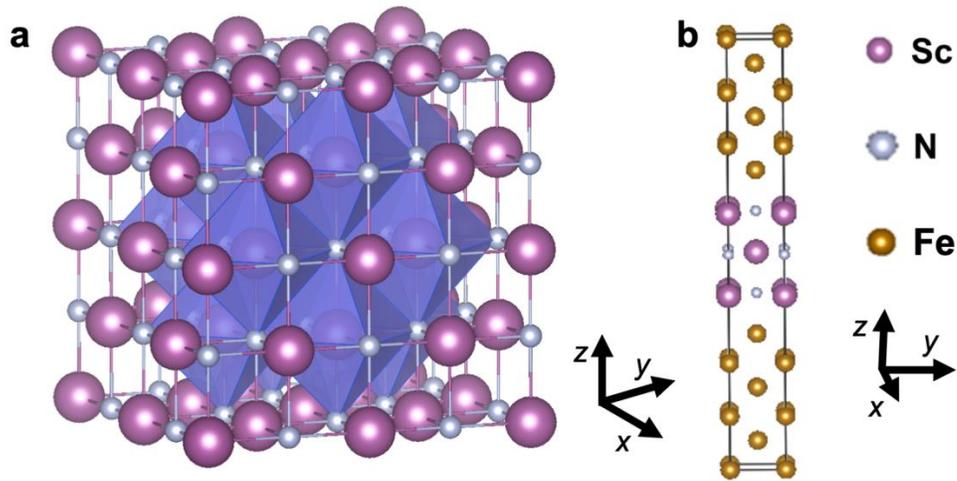

**Figure 1. Crystal structure. a**, Rock salt crystal structure of ScN. **b**, Transport setup of one supercell showing the ScN barrier sandwiched between Fe electrodes with transport along the *z* direction from the bottom Fe to the top Fe layer. Cells are repeated in *x* and *y* to infinity and the Fe electrodes extend to infinity in +*z* and -*z*.

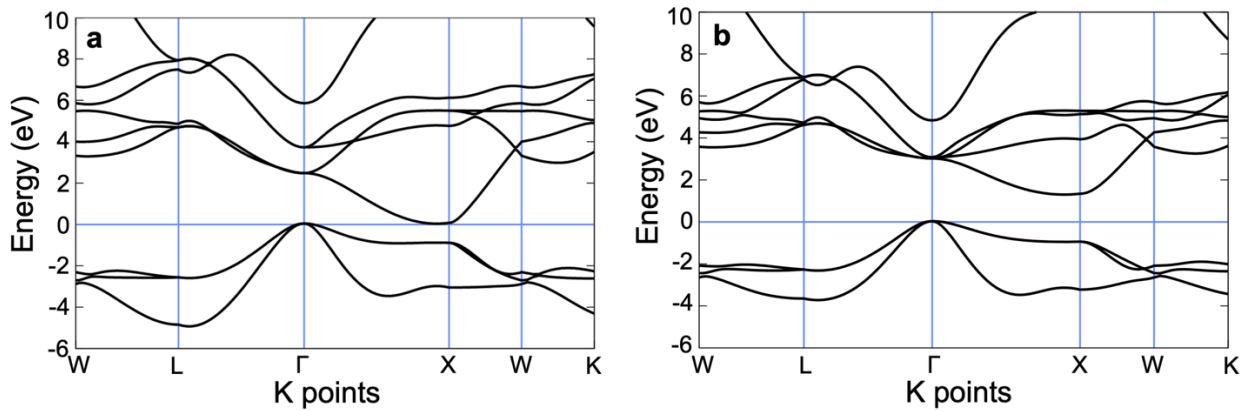

**Figure 2. ScN band structure. a**, Band structure of ScN with direct gap at the Γ point of 2.43 eV. **b**, Band structure of ScN with 4.5 eV Hubbard potential added to the 3d orbital of Sc.

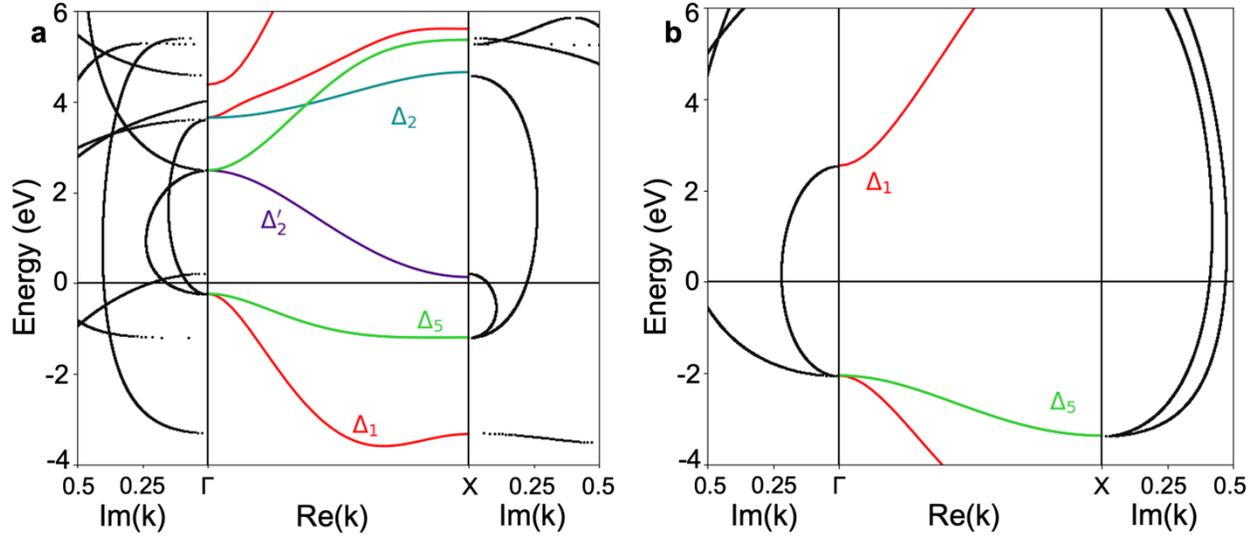

**Figure 3. ScN and MgO complex band structure.** The complex bands of **a**, ScN and **b**, MgO are shown, sampling into imaginary $k$ space from the $\Gamma$ and X points with the symmetry-resolved real bands set in-between.

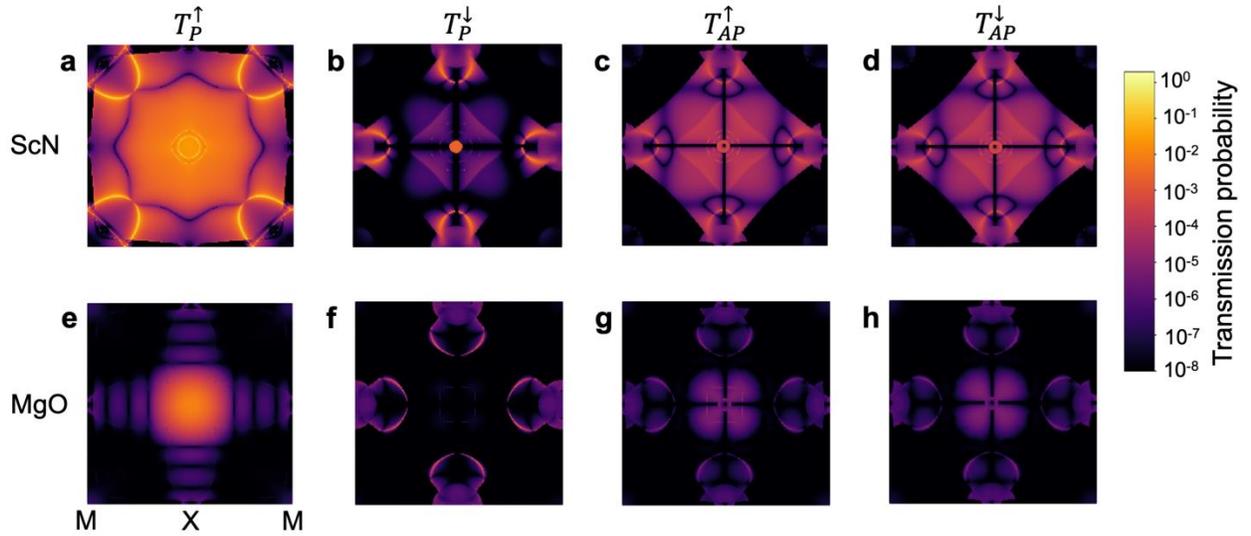

**Figure 4. $k_{||}$-resolved transmission probability. a-d**, Transmission probability through Fe/ScN($t$ = 6)/Fe MTJs, where $t$ is in atomic layers. **e-h**, Transmission probability through Fe/MgO(6)/Fe MTJs. **a,e**, Majority channel for parallel alignment of Fe electrode magnetizations. **b,f**, Minority

channel for parallel alignment of Fe electrode magnetizations. **c,g**, Majority channel (up of input electrode) for antiparallel alignment of Fe electrode magnetizations. **d,h**, Minority channel (down of input electrode) for antiparallel alignment of Fe electrode magnetizations. Color label indicates the transmission probability at each *k* point. Plots are centered on the Γ point with X and M points labeled in **e**.

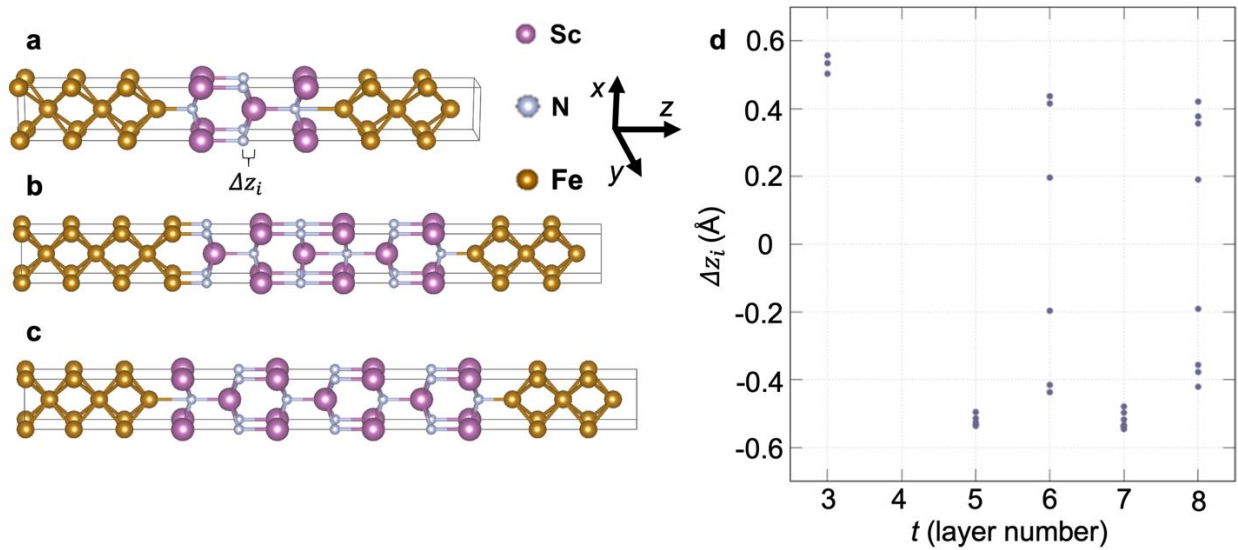

**Figure 5. Sc-N displacement. a**, $t = 3$ layers, **b**, $t = 6$ layers, and **c**, $t = 7$ layers of ScN. Higher *TMR* is observed in the 6 and 8 layer systems where the Sc and N are less displaced from each other in *z*. **d,** The Sc-N interatomic distance $\Delta z_i$ plotted for each atomic layer in every converged system.

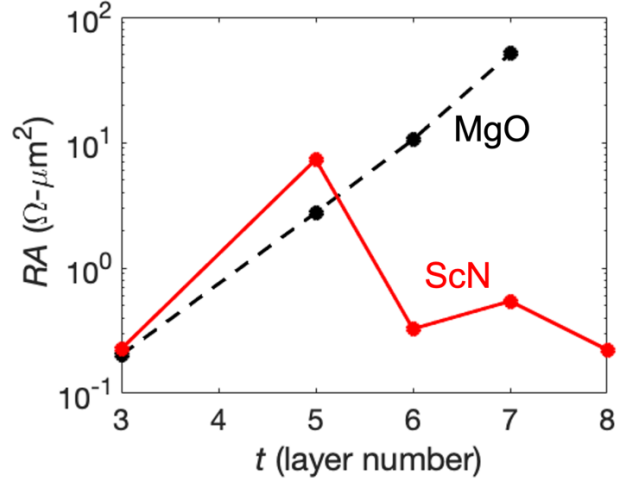

**Figure 6. Resistance-area product.** Comparison of $RA$ for ScN (red, solid) and MgO (black, dotted) MTJs vs. layer number, showing the ScN MTJ $RA$ stays low while the MgO MTJ $RA$ increases with layer number.

**Tables**

| Layers | $G_P^{\uparrow}$ $(e^2/h)$ | $G_P^{\downarrow}$ $(e^2/h)$ | $G_P$ $(\Omega^{-1})$ | $G_{AP}^{\uparrow}$ $(e^2/h)$ | $G_{AP}^{\downarrow}$ $(e^2/h)$ | $G_{AP}$ $(\Omega^{-1})$ | TMR ratio (%) |
|---|---|---|---|---|---|---|---|
| Fe/ScN(3)/Fe | 6.64E-03 | 2.73E-03 | 3.62E-07 | 1.75E-03 | 1.95E-03 | 1.47E-07 | 153 |
| Fe/ScN(5)/Fe | 2.09E-04 | 8.66E-05 | 1.14E-08 | 8.01E-05 | 6.16E-05 | 5.46E-09 | 108 |
| Fe/ScN(6)/Fe | 6.44E-03 | 8.63E-05 | 2.51E-07 | 2.89E-05 | 2.59E-05 | 2.11E-09 | 11,800 |
| Fe/ScN(7)/Fe | 2.96E-03 | 9.62E-04 | 1.51E-07 | 9.63E-05 | 9.84E-05 | 7.51E-09 | 1,920 |
| Fe/ScN(8)/Fe | 9.30E-03 | 2.64E-04 | 3.69E-07 | 5.19E-05 | 8.75E-05 | 5.39E-09 | 6,750 |

**Table 1. ScN MTJ conductance and tunnel magnetoresistance.** For each barrier layer number studied, conductance values for parallel ($G_P$) and antiparallel ($G_{AP}$) alignment in units of ($\Omega^{-1}$) and for majority and minority channels in parallel and antiparallel alignment of the electrodes $G_P^{\uparrow}$, $G_P^{\downarrow}$, $G_{AP}^{\uparrow}$, $G_{AP}^{\downarrow}$, respectively, in units of ($e^2/h$), and corresponding $TMR$.

| Layers | $G_P^\uparrow$ ($e^2/h$) | $G_P^\downarrow$ ($e^2/h$) | ($\Omega^{-1}$) | $G_{AP}^\uparrow$ ($e^2/h$) | $G_{AP}^\downarrow$ ($e^2/h$) | ($\Omega^{-1}$) | TMR ratio (%) |
|---|---|---|---|---|---|---|---|
| Fe/MgO(3)/Fe | 9.88E-03 | 5.94E-04 | 4.04E-07 | 1.12E-03 | 6.82E-04 | 6.96E-08 | 408 |
| Fe/MgO(5)/Fe | 7.55E-04 | 1.34E-05 | 2.96E-08 | 3.88E-06 | 3.81E-06 | 2.96E-10 | 9,700 |
| Fe/MgO(6)/Fe | 1.74E-04 | 2.80E-05 | 7.80E-09 | 3.25E-07 | 2.77E-07 | 2.33E-11 | 30,600 |
| Fe/ MgO(7)/Fe | 4.04E-05 | 1.37E-06 | 1.61E-09 | 6.29E-08 | 3.67E-08 | 3.85E-12 | 41,800 |

**Table 2. MgO MTJ conductance and tunnel magnetoresistance.** For each barrier layer number studied, conductance values for parallel ($G_P$) and antiparallel ($G_{AP}$) alignment in units of ($\Omega^{-1}$) and for majority and minority channels in parallel and antiparallel alignment of the electrodes $G_P^\uparrow$, $G_P^\downarrow$, $G_{AP}^\uparrow$, $G_{AP}^\downarrow$, respectively, in units of ($e^2/h$), and corresponding $TMR$.